\documentclass[12pt]{iopart}
 
\usepackage{epsf} 
\usepackage{setstack}

\begin{document}

\newcommand{\cosech}{\,\mathrm{cosech}}

\title{Large-uncertainty intelligent states for angular momentum and
angle}
\author{J{\"o}rg B G{\"o}tte, Roberta Zambrini, Sonja Franke-Arnold and Stephen M 
  Barnett}
\address{Department of Physics, University of Strathclyde, Glasgow G4 0NG,
  UK}
\ead{joerg@phys.strath.ac.uk}

%%%                                                     
%%% ABSTRACT
%%%

\begin{abstract}
The equality in the uncertainty principle for linear momentum and
position is obtained for states which also minimize the uncertainty
product. However, in the uncertainty relation for angular momentum and angular 
position
both sides of the inequality are state dependent and therefore the 
intelligent states, which 
satisfy the equality, do not necessarily give a minimum for the uncertainty 
product. In this paper, we highlight the difference between intelligent
states and minimum uncertainty states by investigating a class of intelligent
states which obey the equality in the angular uncertainty relation while
having an arbitrarily large uncertainty product. To develop an understanding 
for the uncertainties of angle and angular 
momentum for the large-uncertainty intelligent states we compare exact 
solutions with analytical approximations in 
two limiting cases.
\end{abstract}

\submitto{\JOB}

%%%
%%%
%%%

\maketitle

\newcommand{\erf}{\mathrm{erf}}

%%% 
%%% Introduction
%%%

\section{Introduction}
The uncertainty principle limits the precise knowledge of all physical 
quantities of a system. For linear position and linear momentum this limit is 
expressed as a constant lower bound for the product of the respective 
uncertainties 
\cite{heisenberg+:zphys:1927}:
\begin{equation}
\Delta x \Delta p_x \ge \frac{\hbar}{2}.
\end{equation}
Of a particular interest are states which minimize the uncertainty product, as 
they describe the quantum system as precisely as possible. In the linear case
these states coincide with the intelligent states, that is states, which 
satisfy the equality in the uncertainty relation \cite{arragone}. In the general
case, where the bound in the uncertainty relation is state dependent, one has to 
distinguish between the intelligent states, states which minimize the 
uncertainty product under an additional constraint and states which give a 
global minimum 
\cite{pegg+:njp7:2005,jackiw:jmp9:1968}.

For angular variables like phase and rotation angle it is necessary
to restrict the range of allowed angles to a $2\pi$ radian interval and
to define our observable within this range \cite{phaseop}. Naturally,
the angular uncertainty, the uncertainty relation and the associated
intelligent states will depend on our choice of the $2\pi$ radian range.
Changing the range of angles can greatly modify the angular uncertainty
associated with any given state \cite{methods}. Here, we choose our window
of angles to run from $-\pi$ to $\pi$.  
The uncertainty relation for orbital angular momentum
and angular position has a state dependent lower bound 
\cite{barnettpegg:pra41:1990}:
\begin{equation}
\label{eq:oamur}
\Delta \phi \Delta L_z \ge \frac{\hbar}{2} \bigl| 1 - 2\pi P(\pi) \bigr|.
\end{equation}
as it contains the angle probability density $P(\pi)$ at the angle $\pi$, 
corresponding to the end of our range of angles. 
For a general interval $[\theta_0, \theta_0+2\pi)$, the probability density
in the uncertainty relation (\ref{eq:oamur}) would be evaluated at the 
angle $\theta_0$ or $\theta_0 + 2\pi$, as physical properties are periodic in 
the angle.

The form of the uncertainty relation
(\ref{eq:oamur}) has been confirmed experimentally using light beams carrying
orbital angular momentum \cite{frankearnold+:njp6:2004}.
It is well established that the spin angular momentum of a photon corresponds to
the 
polarisation of a light beam and can be found with one of two values $\pm \hbar$ 
per photon for left- and right-handed circular polarisation. 
The orbital angular momentum, however,
is associated with the azimuthal helical phase fronts $\exp(\rmi m \phi)$ and 
has 
values of $m\hbar$ per photon, where $m$ can take any integer value 
\cite{allen}.
In the experimental verification of the uncertainty relation (\ref{eq:oamur})
a light beam with zero orbital angular momentum was passed through an angular 
aperture, which restricts the angular position and changes the angle 
probability density for the passing light beam \cite{frankearnold+:njp6:2004}. 
The form of the aperture corresponded to the angle probability density of the 
intelligent states. According to the Heisenberg uncertainty
principle a restriction of the angular position leads to a spread in the angular
momentum distribution, which can be measured with various experimental 
techniques 
\cite{oammeasure}. The intelligent states used in the confirmation of the
uncertainty relation had the form of truncated Gaussians in the angle 
representation
\cite{frankearnold+:njp6:2004}
\begin{equation}
\label{eq:truncgauss}
\psi(\phi) = \frac{1}{N} \exp(-\frac{\lambda}{2}\phi^2), \quad -\pi \leq \phi
< \pi \quad \lambda > 0.
\end{equation}
This class of states is characterised by a 
parameter
$\lambda$ and for $\lambda > 0$ the angle wavefunction of the intelligent states
has a single peak at $\phi=0$, which is in the middle of the chosen $2\pi$ radian 
interval. These states and 
their
relation to the constrained minimum uncertainty states have been explored at some 
length \cite{pegg+:njp7:2005,frankearnold+:njp6:2004}.

In this paper we focus on intelligent
states with $\lambda <0$. In contrast to the intelligent states for 
$\lambda > 0$,
whose uncertainty product has an upper bound of $\hbar/2$, intelligent states
with $\lambda < 0$ may have an arbitrarily large uncertainty product and yet still
fulfil the equality in the uncertainty relation (\ref{eq:oamur}) 
\cite{galindopascual:sv:1990}. This is 
because for $\lambda < 0$ the angle wavefunction grows exponentially towards 
the boundaries of the interval $[-\pi,\pi)$ thus increasing the right hand side
of the uncertainty relation.

In the following we will present an analysis of the large-uncertainty 
intelligent states, their
angular momentum distribution and uncertainty product. Along with the 
analytically exact form of the intelligent states, we are able to describe
the limiting behaviour of these states by means of some simple 
analytical approximations. This 
paper complements the analysis of the
angular uncertainty relation started in 
\cite{pegg+:njp7:2005,frankearnold+:njp6:2004}
and emphasizes the difference between intelligent and constrained minimum 
uncertainty product states in the angular case. These are states which minimize
the uncertainty product for the constraint of a given uncertainty in either the
angle or the angular momentum.

%%%
%%% The uncertainty relation
%%%

\section{The uncertainty relation}
\label{sec:ur}
The simplicity of the uncertainty relation (\ref{eq:oamur}) hides a number of 
mathematical subtleties. We can apply it with confidence to any physically
preparable state but its derivation from angle and angular momentum 
observables requires the careful application of a limiting procedure
\cite{barnettpegg:pra41:1990} and we should treat (\ref{eq:oamur}) as
an approximation, albeit an excellent one, to a more rigorous result.
Careless manipulation can lead to difficulties and these have been
ascribed to problems with self-adjointness of the observables \cite{selfadjoint}.
It is clear, however, that such difficulties originate in the inappropriate
application of (\ref{eq:oamur}). In this section we review the formulation
of the angle operator and the derivation of the rigorous uncertainty relation.
This will allow us to determine precisely the limits of validity of
(\ref{eq:oamur}) and of the angle representation of the state as used in this
paper \cite{pegg+:njp7:2005}.

We introduce the angle operator by working with a $(2L+1)$-dimensional 
state space spanned by the eigenvectors $|m\rangle$ of the 
angular momentum operator, $\hat{L}_z$, with $m = -L, -L+1, \dots, -1, 0,
1, \dots, L$. At a later stage, and only after physical results have been
calculated, we shall allow $L$ to tend to infinity. Within the state space
we can construct a complete set of orthonormal angle eigenstates
of the form
\begin{equation}
\label{eq:angleeigen}
|\theta_n\rangle = \left( 2L+1 \right)^{-1/2} \sum_{m=-L}^L
\exp(-\rmi m \theta_n) |m\rangle,
\end{equation}
where
\begin{equation}
\label{eq:angleev}
\theta_n = \theta_0 + \frac{2\pi n}{2L+1}, \quad (n = 0, 1, \dots, 2L).
\end{equation}
The choice of $\theta_0$ is arbitrary and determines the particular basis 
states. In this paper we will use $\theta_0 = -\pi$, so that
the angles lie in the range $-\pi \leq \theta_n < \pi$. The angle
operator $\hat{\phi}_\theta$ has as its eigenstates the states 
$|\theta_n\rangle$
(\ref{eq:angleeigen}) with the associated eigenvalues $\theta_n$ 
(\ref{eq:angleev}):
\begin{equation}
\label{eq:angleop}
\hat{\phi}_\theta = \sum_{n=0}^{2L} \theta_n |\theta_n\rangle \langle \theta_n|.
\end{equation}
This operator is, of course, the natural analogue of the phase operator 
associated
with the phase angle for an optical field mode or harmonic oscillator \cite{phaseop}.
It is manifestly Hermitian and, by virtue of the fact that it
operates on a state space of finite dimension, also self-adjoint. The forms
of the angle operator (\ref{eq:angleop}) and of the angular momentum operator,
\begin{equation}
\hat{L}_z = \hbar \sum_{m=-L}^L m |m\rangle \langle m |,
\end{equation}
lead directly to the commutator
\begin{equation}
\label{eq:commutator}
[\hat{\phi}_\theta, \hat{L}_z] = \frac{2\pi\hbar}{2L+1} 
\sum_{\substack{m, m' /crcr m \neq m'}}
\frac{(m - m') \exp[\rmi(m - m')\theta_0]|m'\rangle \langle m|}
{\exp[\rmi(m - m') 2\pi (2L+1)] - 1}
\end{equation}
and hence to a uncertainty relation of the Robertson-Schr{\"o}dinger form 
\cite{robertson+} 
\begin{equation}
\label{eq:rigoamur}
\Delta L_z \Delta \phi_\theta \geq \frac{1}{2} \bigl| \langle [\hat{\phi}_\theta,
\hat{L}_z] \rangle \bigr|.
\end{equation}
This uncertainty relation is rigorously correct and leads to sensible results. 
In
particular the expectation value of the commutator is zero for both eigenstates
of $\hat{L}_z$ and $\hat{\phi}_\theta$, for which $\Delta L_z=0$ and
$\Delta \phi_\theta = 0$ respectively.

Our simpler form of the uncertainty relation (\ref{eq:oamur}) arises on applying
the limit $L \to \infty$ to the expectation value of the commutator in
(\ref{eq:rigoamur}). \emph{If} the state has finite moments of $\hat{L}_z$ then 
we can
treat $m - m'$ in the denominator of (\ref{eq:commutator}) as small compared 
with
$2L+1$ and this leads to the uncertainty relation of the form
\begin{equation}
\Delta L_z \Delta \phi_\theta \geq \frac{\hbar}{2} \biggl| 1 - (2L+1) 
\biggl\langle|\theta_0\rangle \langle \theta_0| \biggr\rangle \biggr|.
\end{equation}
In the limit as $L \to \infty$, our discrete set of angles approaches a 
continuous 
limit, with angle density $(2L+1)/2\pi$ and we can introduce 
a continuous angle wavefunction $\psi(\phi)$ and 
probability density $P(\phi) = |\psi(\phi)|^2$, normalised such that
\begin{equation}
\label{eq:angleprob}
\int_{\theta_0}^{\theta_0 + 2\pi} P(\phi) \rmd\phi = 1.
\end{equation} 
In this limit our uncertainty relation becomes
\begin{equation}
\label{eq:oamurnod}
\Delta L_z \Delta \phi_\theta \geq \frac{\hbar}{2} | 1 - 2\pi P(\theta_0)|,
\end{equation}
which, on selecting $\theta_0=-\pi$ and using periodicity (which requires 
$P(-\pi) = P(\pi)$),
gives (\ref{eq:oamur}). We emphasize that (\ref{eq:rigoamur}) is the rigorous 
result
and that (\ref{eq:oamurnod}) is an approximation, which is valid for all 
physically 
preparable states. This does not mean, however, that all 
\emph{mathematical operations} using (\ref{eq:oamurnod}) 
and the associated continuous angle variable are 
allowed. Working directly with the continuous variable admits angular momenta
with arbitrarily large values, corresponding to angular Fourier components of
arbitrarily high frequency. This introduces the possibility of angular
wavefunctions with discontinuous derivatives and means that we need to 
exercise caution when identifying $\hat{L}_z$ with $-\rmi \hbar 
(\rmd/\rmd \phi)$
\cite{pegg+:njp7:2005}.  

One cautionary example is provided by considering the intelligent 
states derived in
\cite{pegg+:njp7:2005,frankearnold+:njp6:2004,galindopascual:sv:1990}, for which
we found the (continuous) angle wavefunction
\begin{equation}
\psi(\phi) = \frac{1}{N} \exp(-\frac{\lambda}{2} \phi^2), \quad -\pi \leq 
\phi < \pi.
\end{equation}
This wavefunction has angular momentum amplitudes that fall off like $m^{-2}$  
for large $m$. This means that the action of $\hat{L}_z = -\rmi \hbar 
(\rmd / \rmd \phi)$ on $\psi(\phi)$ leads to a well behaved, that is square 
integrable, state, but the application of $\hat{L}^2_z = 
-\hbar^2 (\rmd^2 / \rmd \phi^2)$ does not. Moreover, we also find a problem 
with the self-adjointness of $-\rmi\hbar (\rmd / \rmd \phi)$ for this 
state in that
\begin{equation}
\int_{-\pi}^\pi \psi^\ast(\phi) \left( -\hbar^2 \frac{\rmd}{\rmd \phi^2} \right)
\psi(\phi) \rmd\phi \neq \hbar^2 \int_{-\pi}^\pi \left| \frac{\rmd \psi}{\rmd 
\phi} \right|^2 \rmd \phi.
\end{equation}
Problems of this kind are, of course, a consequence of performing forbidden 
mathematical operations (associated with the continuous limit) rather than an 
indication of problems with the angle or angular momentum operators.
Such difficulties can always be resolved, by returning to the $2L+1$ 
space and the exact uncertainty relation (\ref{eq:rigoamur}).

%%%
%%% Intelligent states
%%%

\section{Intelligent states}
\begin{figure}
\begin{center}
\epsfbox{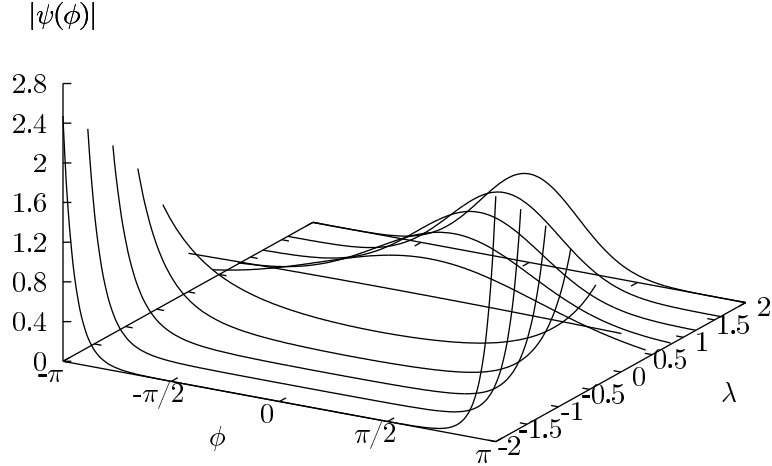}
\caption{\label{fig:wavefunction} The wavefunction in angle representation for 
the 
intelligent states plotted for different values of $\lambda$. The transition 
from 
the truncated Gaussians in \cite{frankearnold+:njp6:2004} to the 
large-uncertainty  intelligent states for 
$\lambda < 0$ is shown.}
\end{center}
\end{figure}

The equality in the general form of the uncertainty relation 
\cite{robertson+} leads to a condition for the intelligent
states \cite{frankearnold+:njp6:2004,schwabl:sv:2002}:
\begin{equation} 
\label{eq:equalcond}
\left[ \hat{L}_z - \langle \hat{L}_z \rangle \right] |\psi\rangle = 
\rmi \hbar \lambda \left[ \hat{\phi}_{\theta} - \langle \hat{\phi}_{\theta} 
\rangle \right] | \psi \rangle,
\end{equation}
where $\lambda$ is a real number. We choose the angle eigenvalues to lie in 
the range $-\pi$ to $\pi$ so that $\theta_0 = -\pi$ and, for simplicity of
notation, denote the corresponding angle operator $\hat{\phi}_{-\pi}$ as
$\hat{\phi}$. It has been shown, moreover, that we can restrict ourselves to
cases with zero mean angle and angular momentum 
\cite{pegg+:njp7:2005,frankearnold+:njp6:2004} without loss of generality.
The equality condition (\ref{eq:equalcond}) can be turned into a
differential equation for the intelligent states by employing the
angle representation $\psi(\phi)$. 
Solutions for the resulting differential equation
\begin{equation}
\left[ \frac{\partial}{\partial \phi} + \lambda \phi \right] 
\psi(\phi) = 0
\end{equation}
for $\lambda > 0$ are the truncated Gaussians (\ref{eq:truncgauss})
\cite{pegg+:njp7:2005,frankearnold+:njp6:2004}. Formally this solution can be 
extended to negative values of $\lambda$ \cite{galindopascual:sv:1990}.
For such negative values the wavefunction in the angle representation is 
given by
\begin{equation}
\label{eq:wavefunction}
\psi(\phi) = \frac{1}{N}\exp(-\frac{\lambda}{2} \phi^2) = \frac{1}{N}\exp(\frac{|\lambda|}{2} \phi^2),
\end{equation}
with the normalisation constant
\begin{equation}
\label{eq:normalisation}
{N^2} = \int_{-\pi}^\pi \exp(|\lambda|\phi^2) \rmd \phi = 
-\rmi \sqrt{\frac{\pi}{|\lambda|}} 
\erf(\rmi \sqrt{|\lambda|}\pi).
\end{equation}
Note that the complex error function with a purely imaginary argument is 
also purely imaginary. For $\lambda < 0$ the wavefunction grows 
exponentially and is inversely proportional to the truncated
Gaussians (see figure (\ref{fig:wavefunction})). A state represented by
this wavefunction can
only exist because it is defined on a finite interval.
For the case of linear momentum and position, where Gaussians are both the
intelligent and constrained minimum uncertainty product states, 
solutions can also be characterised by a parameter $\lambda$. However, for the  
large-uncertainty case with $\lambda < 0$ the wavefunctions would
not be normalisable and therefore not be elements of the Hilbert space.

% Angle uncertainty
\subsection{Angle uncertainty}
\begin{figure}
\begin{center}
\epsfbox{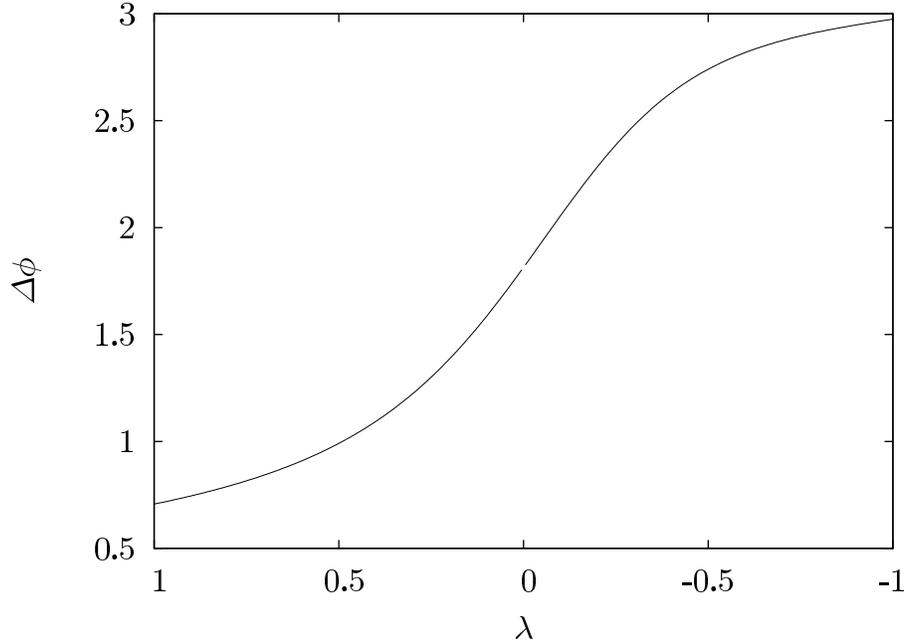}
\caption{\label{fig:deltaphi_ext} The angle uncertainty $\Delta \phi$ for 
the intelligent states. The parameter value $\lambda = 0$ distinguishes 
between
the truncated Gaussians with $\lambda > 0$ \cite{frankearnold+:njp6:2004} and 
the
large-uncertainty intelligent states with $\lambda < 0$. For intelligent states
the angle uncertainty is bounded by $\Delta \phi \leq \pi$.}
\end{center}
\end{figure}

The square of the angle uncertainty is given by $(\Delta \phi)^2 = 
\langle \hat{\phi}^2 \rangle - \langle \hat{\phi} \rangle^2$. As our solutions 
are symmetric about $\phi=0$, the mean value of $\hat{\phi}$ is zero and
the angle uncertainty is simply the expectation value of $\hat{\phi}^2$
\begin{equation}
\label{eq:deltaphi}
\left( \Delta \phi \right)^2 = \int_{-\pi}^\pi \phi^2 P(\phi) \rmd \phi =
\frac{1}{N^2} \int_{-\pi}^\pi \phi^2 \exp(|\lambda| \phi^2) \rmd \phi.
\end{equation}
The functional form of the wavefunction allows us to express the angle 
uncertainty
as the derivative of the normalisation constant $N$ with respect to $|\lambda|$:
\begin{equation}
\label{eq:deltaphidrv}
\left( \Delta \phi \right)^2 = \frac{1}{N^2} \int_{-\pi}^\pi 
\frac{\rmd}{\rmd |\lambda|} \exp(|\lambda|\phi^2) \rmd \phi = 
\frac{1}{N^2} \frac{\rmd N^2}{\rmd |\lambda|}.
\end{equation}
An expression for the angle uncertainty can be
obtained on substituting the analytically exact normalisation
constant (\ref{eq:normalisation}) in (\ref{eq:deltaphidrv}):
\begin{equation}
\label{eq:exctdphi}
\Delta \phi = \left| \rmi \sqrt{\frac{\pi}{|\lambda|}} 
\frac{\exp(|\lambda|\pi^2)} {\erf(\rmi \sqrt{|\lambda|\pi})} - 
\frac{1}{2|\lambda|} \right|^{1/2}.
\end{equation}
The largest angle uncertainty possible for intelligent states
is $\Delta \phi = \pi$. This can be seen from (\ref{eq:deltaphi}) as
$P(\pi)$ is an even function of $\phi$ for the intelligent states: 
\begin{equation}
\left( \Delta \phi \right)^2 = 2 \int_0^\pi \phi^2 P(\phi) \rmd \phi 
\leq 2 \pi^2 \int_0^\pi P(\phi) \rmd \phi = \pi^2.
\end{equation}
The maximum value $\Delta \phi = \pi$ is obtained in the limit 
of $\lambda \to -\infty$. A plot of $\Delta \phi$ as a function
of $\lambda$ is shown in figure \ref{fig:deltaphi_ext}.
For $\lambda = 0$ the flat angle probability density $P(\phi) = 1/(2\pi)$ 
gives the angle uncertainty of $\Delta \phi = \pi/\sqrt{3}$. The truncated
Gaussians with $\lambda > 0$ have a smaller angle uncertainty than 
$\pi/\sqrt{3}$ while the intelligent states with $\lambda < 0$ considered 
here have a larger angle uncertainty.

% Angular momentum uncertainty  
 \subsection{Angular momentum uncertainty}
\begin{figure}
\begin{center}
\epsfbox{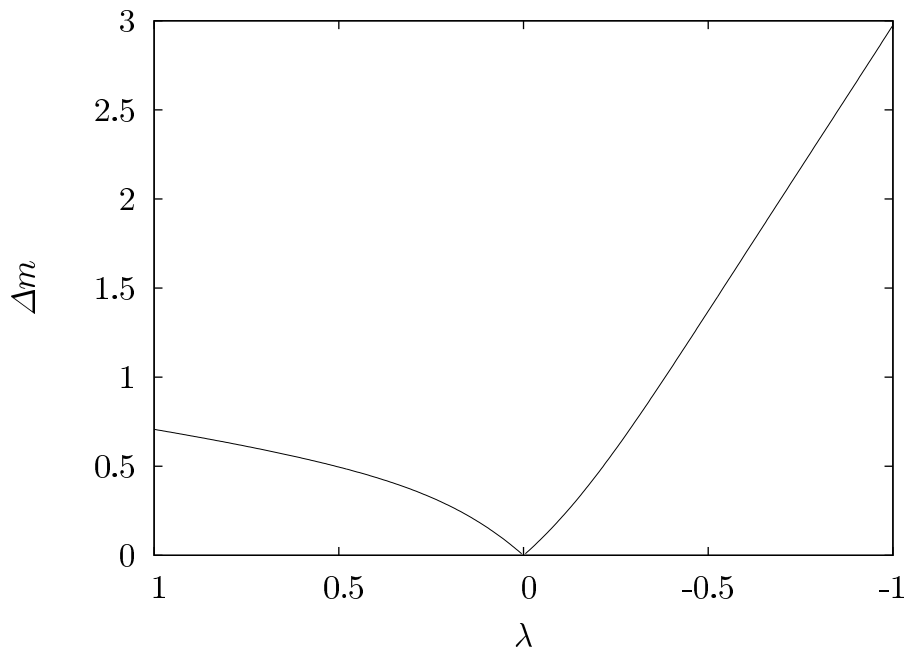}
\caption{\label{fig:deltam_ext} The angular momentum uncertainty 
$\Delta m$ plotted as a function of $\lambda$. From 
(\ref{eq:deltam}), which also holds for $\lambda > 0$, one can see
that $\Delta m$ tends to infinity for $\lambda \to \infty$. For 
$\lambda \to -\infty$, $\Delta m$ tends also to infinity and for this
case the behaviour is visible in the plot.}
\end{center}
\end{figure}

Using the continuous wavefunction (\ref{eq:wavefunction}) and
the representation of $\hat{L}_z$ as a derivative requires
some care. In section \ref{sec:ur} we have argued that
for the intelligent states
the first derivative of $\psi(\phi)$ with respect to $\phi$
represents a well behaved state, while the second derivative
does not. This is the reason why we can use $-\rmi \hbar (\rmd/\rmd \phi)$
as a valid representation of $\hat{L}_z$ in the eigenvalue equation 
for the intelligent states.
However, in the arbitrarily large state space of $2L+1$ dimension reviewed in
section \ref{sec:ur} the angular momentum operator $\hat{L}_z$ is 
self-adjoint and Hermitian, and we may therefore write for the
expectation value of $\hat{L}_z^2$ for the intelligent state
$|\psi\rangle$:
\begin{equation}
\langle \psi | \hat{L}_z^2 \psi \rangle = 
\langle \hat{L}_z \psi | \hat{L}_z \psi \rangle = 
\|\, |\hat{L}_z \psi \rangle\, \|^2,
\end{equation}
where $\| \cdot \|$ symbolises the norm of a state vector.
The expectation value is a physical quantity and we can make now
the transition to the continuous wavefunction $\psi(\phi) = \langle \phi
| \psi \rangle$. Following the argument given above it is then
valid to replace $\hat{L}_z$ with $-\rmi \hbar (\rmd/\rmd \phi)$, allowing
$(\Delta m)^2$ to be written as
\begin{equation}
( \Delta m)^2 = \frac{1}{\hbar^2} \langle \hat{L}^2_z \rangle
= \frac{1}{N^2} \int_{-\pi}^\pi \left| \frac{\rmd}{\rmd \phi} \psi(\phi)
\right|^2 \rmd \phi.
\end{equation}
On substituting the wavefunction (\ref{eq:wavefunction}) into this
expression we find that the angular momentum uncertainty $\Delta m$ 
can be expressed
in terms of the angle uncertainty $\Delta \phi$:
\begin{equation}
\label{eq:deltam}
( \Delta m)^2 =  \frac{1}{N^2} \int_{-\pi}^\pi |\lambda|^2 \phi^2 
\exp(|\lambda| \phi^2) \rmd \phi.\nonumber = |\lambda|^2 (\Delta \phi)^2.
\end{equation}
If we use the expression for the angle uncertainty $\Delta \phi$ 
from (\ref{eq:exctdphi}) we obtain for the angular momentum
uncertainty $\Delta m$:
\begin{equation}
\label{eq:deltam_ext}
\Delta m = |\lambda| \left| \rmi \sqrt{\frac{\pi}{|\lambda|}} 
\frac{\exp(|\lambda|\pi^2)} {\erf(\rmi \sqrt{|\lambda|\pi})} - 
\frac{1}{2|\lambda|} \right|^{1/2}.
\end{equation}
To obtain the equality in the uncertainty relation for linear position
and momentum, the uncertainties have to be inversely proportional
to each other. In the angular case the uncertainties are 
directly proportional to each other; if we have a large 
angle uncertainty we require also a large angular momentum
uncertainty to satisfy the equality. The reason for this
behaviour lies in the state dependent right hand side
of (\ref{eq:oamur}). A plot of the angular momentum 
uncertainty for positive and negative values of $\lambda$
is shown in figure (\ref{fig:deltam_ext}).
For $\lambda=0$ we have the flat angle probability distribution and
an angular momentum eigenstate $|m\rangle$ with zero angular momentum
uncertainty $\Delta m = 0$. In our case, with zero 
angular momentum mean, we have $m=0$ and the eigenstate $|0\rangle$.

% Uncertainty product 
\subsection{Uncertainty product}
\begin{figure}
\begin{center}
\epsfbox{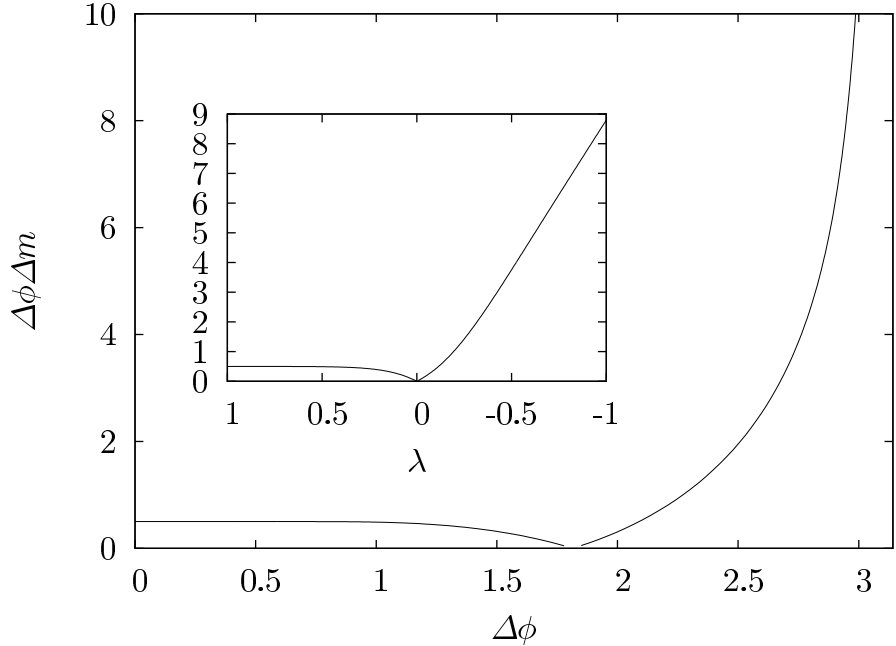}
\caption{\label{fig:uncertaintyprod_ext} Plot of the uncertainty product 
$\Delta m \Delta\phi$ against $\Delta \phi$. For the large-uncertainty
intelligent states ($\Delta \phi > \pi/\sqrt{3}$ or $\lambda < 0$) the
uncertainty product has no upper bound. The inset shows a plot of the
uncertainty product against $\lambda$ for comparison with the plots
of $\Delta \phi$ and $\Delta m$ (see figures \ref{fig:deltaphi_ext} and
\ref{fig:deltam_ext}; see also a similar plot in 
\cite{galindopascual:sv:1990})}
\end{center}
\end{figure}

The left hand side of the uncertainty relation (\ref{eq:oamur}) 
is given by the uncertainty product $\Delta m \Delta \phi$. Using
the result for the angular momentum uncertainty $\Delta m$ 
(\ref{eq:deltam}) and the expression for $\Delta \phi$ 
(\ref{eq:exctdphi}) the uncertainty product can be expressed as 
\begin{equation}
\label{eq:oamur_lhs}
\Delta m \Delta \phi = |\lambda| (\Delta \phi)^2 = 
\rmi \sqrt{\pi|\lambda|} 
\frac{\exp(|\lambda|\pi^2)} {\erf(\rmi \sqrt{|\lambda|\pi})} - 
\frac{1}{2}
\end{equation}
The right hand side of the uncertainty relation is given by
$|1-2\pi P(\pi)|/2$. For the large-uncertainty intelligent
states with $\lambda < 0$, $P(\pi)$ is always larger than 
$1/(2\pi)$. The modulus can thus be replaced by 
$2\pi P(\pi) - 1$. Using the wavefunction (\ref{eq:wavefunction})
and the normalisation (\ref{eq:normalisation}) to express
$P(\pi)$ yields the equality with the uncertainty product.
The behaviour of the
uncertainty product as a function of $\Delta \phi$ is shown in
figure \ref{fig:uncertaintyprod_ext}. For $\Delta \phi < \pi/\sqrt{3}$
this is the same plot as in \cite{frankearnold+:njp6:2004}. From figure
\ref{fig:uncertaintyprod_ext} and equation (\ref{eq:oamur_lhs}) it is clear 
that for the intelligent states 
with $\lambda < 0$ the uncertainty product can become arbitrarily large.

%%% Angular momentum distribution
\subsection{Angular momentum distribution}
\begin{figure}
\begin{center}
\epsfxsize=0.49\textwidth
\epsfbox{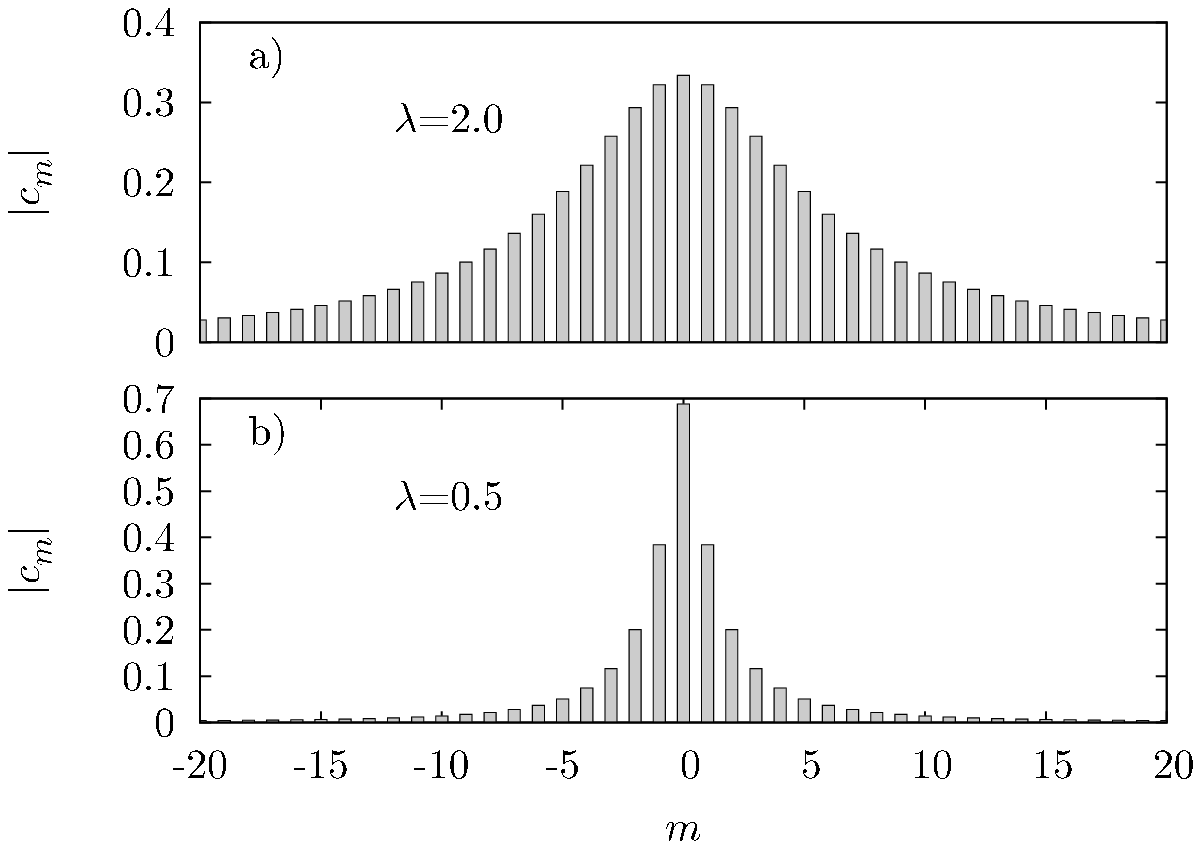}
\epsfxsize=0.49\textwidth
\epsfbox{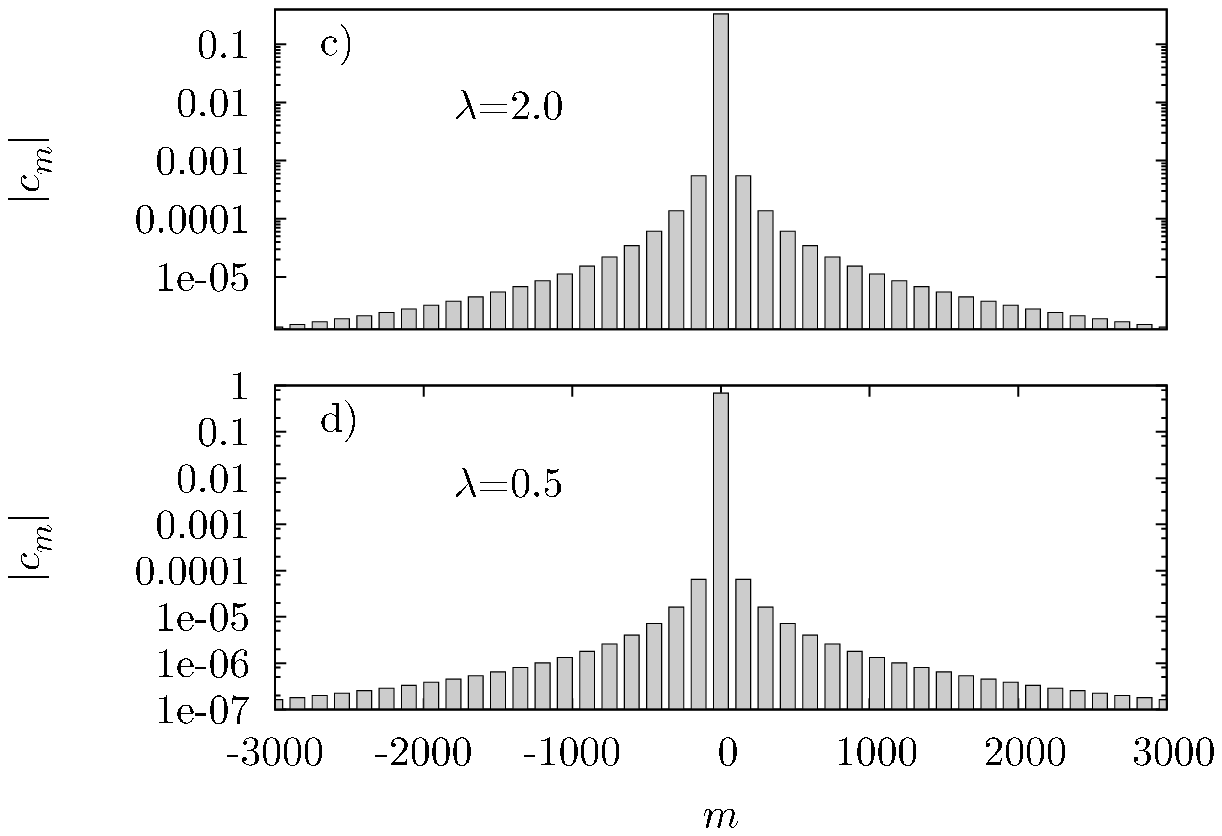}
\caption{\label{fig:oamdist} Orbital angular momentum distribution for different 
values of $\lambda$. a) $\lambda=-2.0$: A linear plot showing the distribution
for the central $m$ values. b) $\lambda=-0.5$: linear plot. 
c) $\lambda=-2.0$: A logarithmic
plot for the flanks. The width of the bars in the plot covers 100 $m$ values,
but the shown value corresponds to the $m$ value in the middle of the bar.
d) $\lambda=-0.5$: logarithmic plot.}
\end{center}
\end{figure}

The probability amplitudes of the orbital angular momentum $c_m$ are 
calculated from the wavefunction (\ref{eq:wavefunction}) by means of a Fourier
transform:
\begin{equation}
c_m = \frac{1}{\sqrt{2\pi}}\int_{-\pi}^\pi \psi(\phi) \exp(-\rmi m \phi)
\rmd \phi.
\end{equation}
By virtue of the symmetry of the wave function the transformation
may be specialised to the Fourier cosine transform.
The result can be expressed in terms of a complex error
function
\begin{equation}
\label{eq:oamprobampl}
c_m =\frac{(-1)^m}{N\sqrt{2\pi}} \exp\left(\frac{m^2}{2|\lambda|}\right) 
\sqrt{\frac{\pi}{2|\lambda|}} 2 \: \Im \: 
\left[ \erf\left( \sqrt{\frac{1}{2|\lambda|}}m + 
\rmi\sqrt{\frac{|\lambda|}{2}} \pi \right) \right],
\end{equation}
where $\Im$ denotes the imaginary part of a complex number. The divergent 
behaviour of the positive exponent is counterbalanced
by the imaginary part of the complex error function, which decays
quickly to zero in this case. 
This results in an angular momentum distribution which is
similar to a Lorentzian (see figure (\ref{fig:oamdist})).

%%%
%%% Limiting behaviour
%%%

\section{Limiting behaviour}
In the following we investigate the large-uncertainty intelligent states
for two limiting cases to develop an understanding for the
behaviour of the angle and angular momentum uncertainties. For small
$|\lambda|$ we present a perturbative approach, while for large
$\lambda$ we approximate the exponential in the 
wavefunction (\ref{eq:wavefunction}).

% Small lambda
\subsection{Small $|\lambda|$ approximation}
\begin{figure}
\begin{center}
\epsfbox{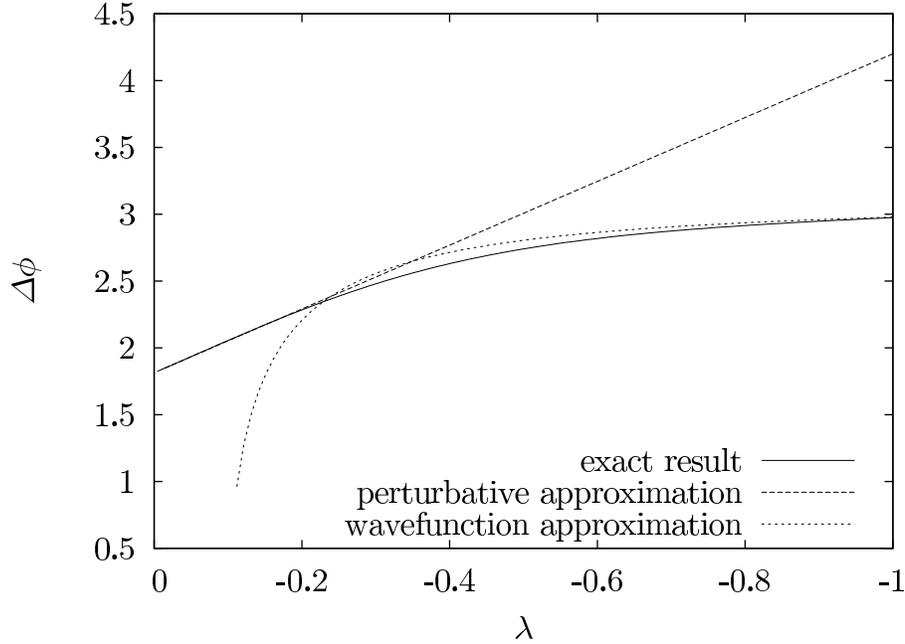}
\caption{\label{fig:deltaphi} The angle uncertainty $\Delta \phi$ plotted
for the analytically exact expression (\ref{eq:exctdphi}), in the perturbative
approximation and in the wavefunction approximation (\ref{eq:wvfdphi}). The wavefunction
approximation fails for small values of $|\lambda|$. In this region a
perturbative approximation can explain the behaviour of $\Delta \phi$.}
\end{center}
\end{figure}

\begin{figure}
\begin{center}
\epsfbox{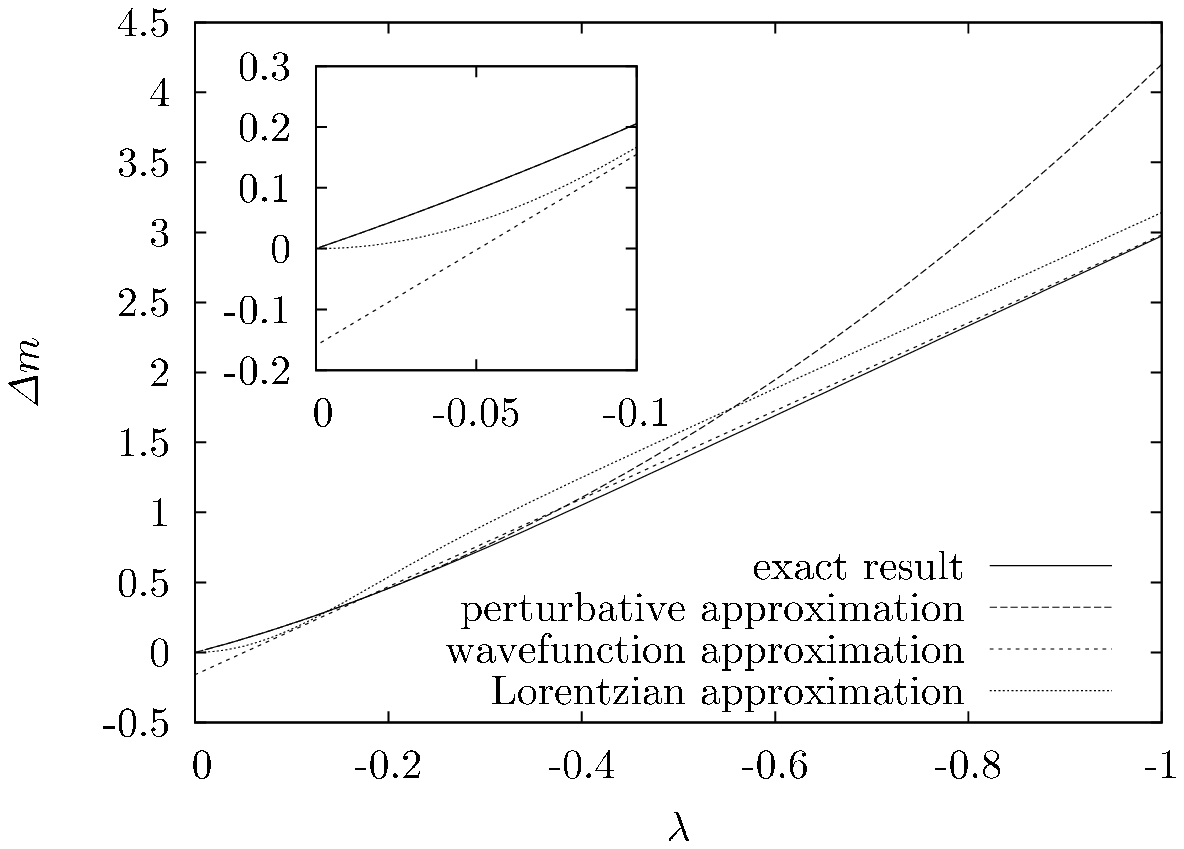}
\caption{\label{fig:oamuncertainty} Plot of the angular momentum uncertainty
for the exact result and different approximations. The expressions for 
$\Delta m$ in the wavefunction and perturbative approximation is derived 
from (\ref{eq:deltam}) by substituting the corresponding expression for
$\Delta \phi$. For very small $|\lambda|$ the perturbative 
approximation is indistinguishable from the exact result.
The Lorentzian approximation originates from a summation of
the Lorentzian angular momentum probabilities in (\ref{eq:lordm}).}
\end{center}
\end{figure}

For small values of $|\lambda|$ the continuous
angle probability density becomes a flat function of the angle and in the 
limit of $\lambda \to 0$, where $P(\phi) = 1/(2\pi)$ we have an angular 
momentum eigenstate. As we consider the case with zero angular momentum mean
$\langle \hat{L}_z \rangle=0$, the angular momentum eigenstate in the limit 
of $\lambda \to 0$ is $|0\rangle$. The behaviour of $\Delta \phi$ in 
this 
parameter region can be explained using a perturbation ansatz for $|0\rangle$ 
with
$\lambda$ as perturbation parameter. As explained in section 
\ref{sec:ur} we start the derivation of $\Delta \phi$ in this perturbative
approach within an arbitrarily large, but finite state space of
$2L+1$ dimension. For zero angle and angular momentum mean
the condition for the intelligent state (\ref{eq:equalcond}) can be written
as
\begin{equation}
\label{eq:pertequal}
\hat{L}_z |\psi\rangle - \rmi \lambda \hat{\phi} | \psi \rangle = 0.
\end{equation}
For small $\lambda$ we can use the perturbation ansatz $|\psi\rangle = 
|0\rangle + \lambda | \varphi^1 \rangle$. Substituting this
ansatz in the condition (\ref{eq:pertequal}) yields at first order
in $\lambda$
\begin{equation}
\label{eq:linequal}
\lambda \left( \hat{L}_z |\varphi^1\rangle - \rmi \hat{\phi} | 0 \rangle
\right) = 0.
\end{equation}
Without loss of generality we can write the perturbative state $|\varphi\rangle$ 
as a superposition of angular momentum eigenstates $|m\rangle, 
m=-L,-L+1,\dots,-1,1,\dots,L$ without a contribution from $|0\rangle$:
\begin{equation}
|\varphi^1\rangle = \sum_{\substack{m=-L /crcr m \neq 0}}^L c_m^1 |m\rangle.
\end{equation}
The angular momentum eigenstate $|0\rangle$ is a physical state, that
is a state which may be approximated to any desired accuracy by the
expansion $\sum_m b_m |m\rangle$, where the coefficients $b_m =0$ for 
$|m| > M$. Here, the sum includes all integer values of $m$ and the bound $M$ is 
sufficiently large to guarantee the desired
accuracy but always less than $L$. Restricting the domain of the
angle operator to these physical states simplifies the expression
for the angle operator and changes the summation to include an infinite
number of angular momentum states $|m\rangle$.
Using the definitions for the angular momentum operator and for the angle
operator for physical states
in \cite{barnettpegg:pra41:1990} we can calculate the resulting
states $\hat{L}_z |\varphi^1\rangle$ and $\hat{\phi}|0\rangle$:
\begin{eqnarray}
\hat{L}_z |\varphi^1 \rangle & = &  \sum_{\substack{m=-L /crcr m \neq 0}}^L m 
c_m^1 |m\rangle = \sum_{m=-L}^L m  c_m^1 |m\rangle, \\
\hat{\phi}| 0 \rangle & = & -\rmi \sum_{m \neq 0}
\frac{\exp(\rmi m \pi)}{m} |m\rangle. 
\end{eqnarray}
On substituting these results in the condition for the intelligent states
at first order in $\lambda$ (\ref{eq:linequal}) we find an 
equation to determine the $c_m^1$:
\begin{equation}
\lambda \sum_{m \neq 0} \left( m c_m^1 - \frac{\exp(\rmi m \pi)}{m} \right)
|m\rangle = 0.
\end{equation}
For the linearly independent 
basis sates $|m\rangle$ the coefficients in (\ref{eq:linequal}) have
to vanish to give zero as result.
This requires $c_m^1 = -(-1)^m / m^2$. For small $\lambda$ the 
perturbative state 
$|\psi\rangle$ may thus be written as
\begin{equation}
|\psi\rangle = \frac{1}{N_{\mathrm{per}}} \left( |0\rangle - 
\lambda \sum_{m \neq 0} \frac{(-1)^m}{m^2}
|m\rangle \right),
\end{equation}
where $N_{\mathrm{per}}$ is the normalisation
constant for the perturbative wavefunction.
The angular momentum amplitudes of the state $|\psi\rangle$ are given
by $c_0=1/N_{\mathrm{per}}$ and $c_m = \lambda c_m^1/N_{\mathrm{per}}$. 
The angle representation of this state is obtained by projection on an
angle eigenstate. We hereby calculate a physical quantity, the angle
probability amplitude, and we can now allow $L$ to tend to infinity, which
yields the continuous wavefunction of the perturbative state:
\begin{eqnarray}
\label{eq:pertwavefunc}
\langle \phi | \psi \rangle = \psi(\phi) & = & \frac{1}{N_{\mathrm{per}}}
\frac{1}{\sqrt{2\pi}} \left( 1 - \lambda \sum_{m \neq 0}  
\frac{(-1)^m}{m^2} \exp(\rmi m \phi) \right) \nonumber \\
& = & \frac{1}{N_{\mathrm{per}}}
\frac{1}{\sqrt{2\pi}} \left[ 1 + \lambda \left( \frac{\pi^2}{6} - 
\frac{\phi^2}{2} \right) \right].
\end{eqnarray}
The sum in the last equation has been evaluated using the contour integration
method \cite{stephenson+:cup:1993}, which can also be used to calculate
the normalisation from the requirement that the angular momentum probabilities
$|c_m|^2$ sum to unity:
\begin{equation}
\label{eq:pertnorm}
\sum_m |c_m|^2 = 1 = \left( 1 + \lambda^2 
\sum_{m \neq 0} \frac{1}{m^4} \right) =
\left( 1 + \lambda^2 \frac{\pi^4}{45}
\right) \frac{1}{N_{\mathrm{per}}^2}.
\end{equation}
The square of the uncertainty for the perturbative state can be
calculated at first order from the continuous wavefunction 
(\ref{eq:pertwavefunc}):
\begin{equation}
(\Delta \phi)^2 = \langle \hat{\phi}^2 \rangle =
\frac{1}{N_{\mathrm{per}}^2} \frac{1}{2\pi} \int_{-\pi}^\pi
\phi^2 \left[ 1 + \lambda \left( \frac{\pi^2}{6} - \frac{\phi^2}{2}\right) 
\right] \rmd \phi.
\end{equation}
At first order in $\lambda$ this leads to the angle uncertainty
\begin{equation}
\label{eq:pertdphi}
\Delta \phi \approx \frac{\pi}{\sqrt{3}} \left( 1 - \frac{2}{15} \lambda
\pi^2 \right),
\end{equation}
which explains the behaviour of $\Delta \phi$ for small $\lambda$ (see
figure \ref{fig:deltaphi}).
The angular momentum uncertainty can be derived from the angle uncertainty
according to (\ref{eq:deltam}).
Because the angle uncertainty is multiplied by $|\lambda|$
and the perturbative $\Delta \phi$ from (\ref{eq:pertdphi}) is 
at first order $\lambda$ this approximation holds for the second order
in $\lambda$
\begin{equation}
\label{eq:pertdm}
\Delta m \approx |\lambda| \frac{\pi}{\sqrt{3}} \left( 1 -
\frac{2}{15} \lambda \pi^2 \right).
\end{equation}
The behaviour of the angular momentum uncertainty in the perturbative
approach is shown in figure \ref{fig:oamuncertainty}. The equality in 
the uncertainty relation (\ref{eq:oamur}) at first order in  
$\lambda$ can be seen directly from the results for 
$\Delta \phi$ (\ref{eq:pertdphi}) and $\Delta m$ (\ref{eq:pertdm}).
The uncertainty product is given by
\begin{equation}
\Delta m \Delta\phi = |\lambda| \frac{\pi^2}{3} + \Or(|\lambda|^2)
\end{equation}
After substituting the expression for $N_{\mathrm{per}}$ from
(\ref{eq:pertnorm}) into the wavefunction (\ref{eq:pertwavefunc}),
the right hand side of the uncertainty relation becomes
\begin{equation}
\frac{1}{2}\left| 1 - 2\pi P(\pi) \right|
\frac{1}{2} \left| 1 - \left( 1 - 2|\lambda|\frac{\pi^2}{3} + 
\Or(|\lambda|^2) \right) \right| = |\lambda| \frac{\pi^2}{3} + 
\Or(|\lambda|^2)
\end{equation}
which shows the equality in the uncertainty relation to the first order
in $|\lambda|$. The uncertainty product as a function of $\Delta \phi$ 
is shown in figure \ref{fig:uncertaintyprod}.

% Large lambda
\subsection{Large $|\lambda|$ approximation}
In the approximation for large $|\lambda|$ we approximate the 
wavefunction within an integration. This leads to two variants of this
approximation: approximating the wavefunction in the normalisation integral gives a
simple elementary expression, which can be used to calculate
the angle and angular momentum uncertainties and hence the
uncertainty product. We will refer to this approach as wavefunction 
approximation. On the other hand, applying this approximation
under a Fourier integral to calculate the angular momentum
probability amplitudes, allows us to explain the angular momentum
distribution and to calculate the angular momentum uncertainty in this
approximation. This approach will be called Lorentzian approximation as
it explains the Lorentzian shape of the angular momentum distribution.

% Wavefunction approximation
\subsubsection{Wavefunction approximation}
The interesting quantities for the uncertainty relation are all
related to the normalisation constant: $\Delta m = |\lambda| \Delta \phi$
by virtue of (\ref{eq:deltam}) and $\Delta \phi$ is related to the
normalisation according to (\ref{eq:deltaphidrv}). An approximation for 
the normalisation constant in the limit of large $|\lambda|$ provides
simple approximate expressions, which explain the behaviour of 
the uncertainties in this limit.

The normalisation integral (\ref{eq:normalisation}) can be rewritten as
\begin{equation}
N^2 = 2 \exp(|\lambda|\pi^2) \int_{-\pi}^0 \exp[|\lambda|(\phi+\pi)(\phi-\pi)] \rmd\phi.
\end{equation}
For a large $|\lambda|$ only a small region around $-\pi$ will contribute
significantly to the integral, and we can therefore approximate the
factor $(\phi-\pi)$ in the exponential with $-2\pi$ and extend the upper
integration boundary to infinity. This results in the normalisation
constant $N_{\mathrm{wvf}}$
\begin{equation}
\label{eq:wvfnorm}
N_{\mathrm{wvf}}^2 = 2 \exp(|\lambda|\pi^2) \int_{-\pi}^\infty
\exp(-2\pi|\lambda|(\phi+\pi)) \rmd \phi = \frac{1}{|\lambda|\pi}
\exp(|\lambda|\pi^2).
\end{equation}
Using this normalisation constant (\ref{eq:wvfnorm}) in the expression for
the probability density yields $P(\pi)$ this approximation.
From (\ref{eq:wavefunction}) we have:
\begin{equation}
\label{eq:wvfphipi}
P(\pi) \approx |\lambda|\pi \quad \mathrm{for}\:\,|\lambda|\pi^2 > 3.
\end{equation}
We thus can give an approximation of the right hand side of the uncertainty 
relation (\ref{eq:oamur}) in the limit of large $|\lambda|$. The angle and
angular momentum uncertainties in the left hand side can be obtained using
(\ref{eq:deltaphidrv}) and (\ref{eq:deltam}).
Substituting 
$N_{\mathrm{wvf}}$ into (\ref{eq:deltaphidrv}) yields a simple
expression for the angle uncertainty:
\begin{equation}
\label{eq:wvfdphi}
\Delta \phi \approx \frac{|\lambda|\pi}{\exp(|\lambda|\pi^2)} 
\frac{\rmd}{\rmd |\lambda|} \frac{\exp(|\lambda|\pi^2)}{|\lambda|\pi} =
\pi^2 - \frac{1}{|\lambda|}.
\end{equation}
A graph of the angle uncertainty is given in figure \ref{fig:deltaphi}.
One can see that the wavefunction approximation gives a good agreement
with the analytical values for $|\lambda| > 1/3$ but 
fails for smaller values, where $1/|\lambda|$ grows without bound.  
The angular momentum uncertainty in this approximation follows from 
substituting $\Delta \phi$ (\ref{eq:wvfdphi}) in (\ref{eq:deltam})
\begin{equation}
\label{eq:wvfdm}
\Delta m \approx |\lambda|\pi \sqrt{1 - \frac{1}{|\lambda|\pi^2}} 
\approx |\lambda|\pi - \frac{1}{2\pi}.
\end{equation}
This approximate expression for $\Delta m$ gives a good approximation
for large $|\lambda|$ as can be seen in figure \ref{fig:oamuncertainty}.

\begin{figure}
\begin{center}
\epsfbox{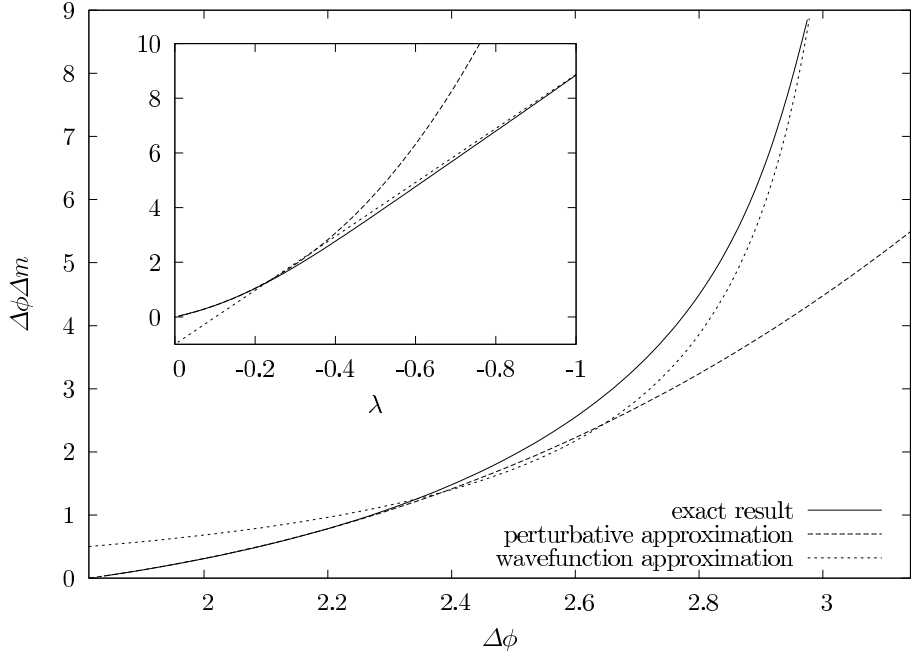}
\caption{\label{fig:uncertaintyprod} Plot of the uncertainty product
$\Delta m \Delta \phi$ as a function of $\Delta \phi$. For negative $\lambda$
$\Delta \phi$ ranges between $\pi/\sqrt{3}$ and $\pi$. The curves are for
the kernel wavefunction approximation for large $|\lambda|$
and for the perturbative approximation for small $|\lambda|$. 
The inset shows the uncertainty product plotted as a function of $\lambda$.}
\end{center}
\end{figure}

In the approximation for large values of $|\lambda|$ the uncertainty product
$\Delta m \Delta \phi$ is given as a simple function of $|\lambda|$. From
(\ref{eq:wvfdphi}) and (\ref{eq:wvfdm}) follows:
\begin{equation}
\label{eq:wvfprod}
\Delta m \Delta \phi \approx |\lambda| \pi^2 - 1.
\end{equation}
The right hand side of the uncertainty relation (\ref{eq:oamur}) contains
the probability density $P(\pi)$ which is approximately $|\lambda|\pi$
from (\ref{eq:wvfphipi}) for $|\lambda|\pi^2| > 3$. Therefore, 
$2\pi P(\pi) = 2|\lambda|\pi^2 > 1$, and the equality in the uncertainty 
relation is expressed by
\begin{equation}
|\lambda| \pi^2 - 1 \simeq |\lambda| \pi^2 - \frac{1}{2} \quad
\mathrm{for}\:\,|\lambda|\pi^2 > 3. 
\end{equation}
Clearly, for sufficiently large values of $|\lambda|$ the discrepancy becomes
negligible. The approximation for large $\lambda$ is based on our 
expression for $N_{\mathrm{wvf}}$. If we refine the approximation 
in (\ref{eq:wvfnorm}) by setting $\exp[|\lambda|(\phi+\pi)(\phi-\pi)] =
\exp[-2\pi|\lambda|(\phi+\pi)]\exp[|\lambda|(\phi-\pi)^2]$ and expanding
the second exponential according to $\exp[|\lambda|(\phi-\pi)^2] \approx
1 + |\lambda|(\phi-\pi)^2$, we find that the expression for $P(\pi)$
changes to
\begin{equation}
P(\pi) = \lambda\pi \left( 1 - \frac{1}{2}\frac{1}{\pi^2|\lambda|}
\right)^{-1}.
\end{equation}
Substituting this expression in the right hand side of the uncertainty
relation (\ref{eq:oamur}) gives the equality with the uncertainty product
(\ref{eq:wvfprod}). A plot of the uncertainty product is given in
figure \ref{fig:uncertaintyprod}.
Compared to the good agreement with the numerical values for the
separate uncertainties $\Delta \phi$ and $\Delta m$, the discrepancies
for the product appear to be large for the wavefunction approximation in
figure \ref{fig:uncertaintyprod}.
This is because  $\Delta m = \lambda \Delta \phi$ [cf. equation (\ref{eq:deltam})]
and as we have plotted the uncertainty product against $\Delta \phi$, the effect of 
an 
error in our approximation for $\Delta \phi$ is enhanced in this plot.

 % Lorentzian approximation
\subsection{Lorentzian approximation}
\begin{figure}
\begin{center}
\epsfxsize=0.49\textwidth
\epsfbox{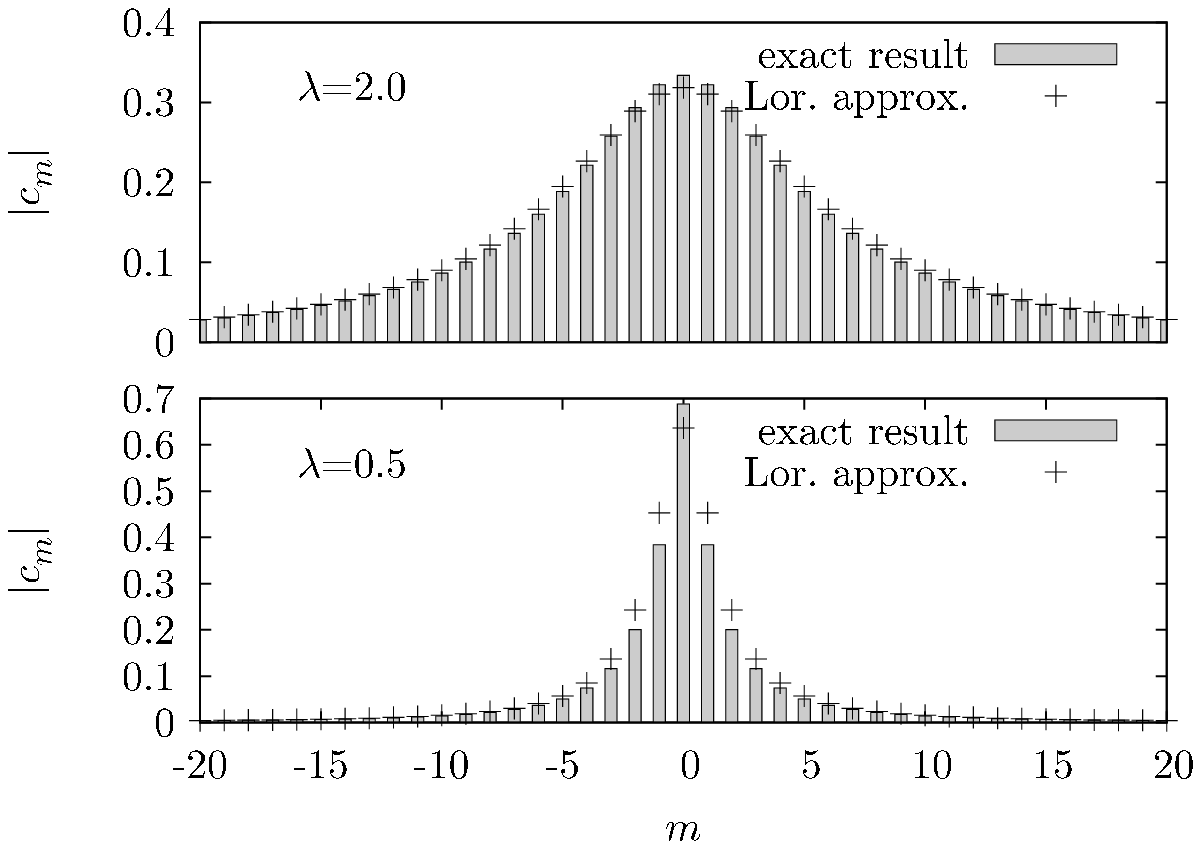}
\epsfxsize=0.49\textwidth
\epsfbox{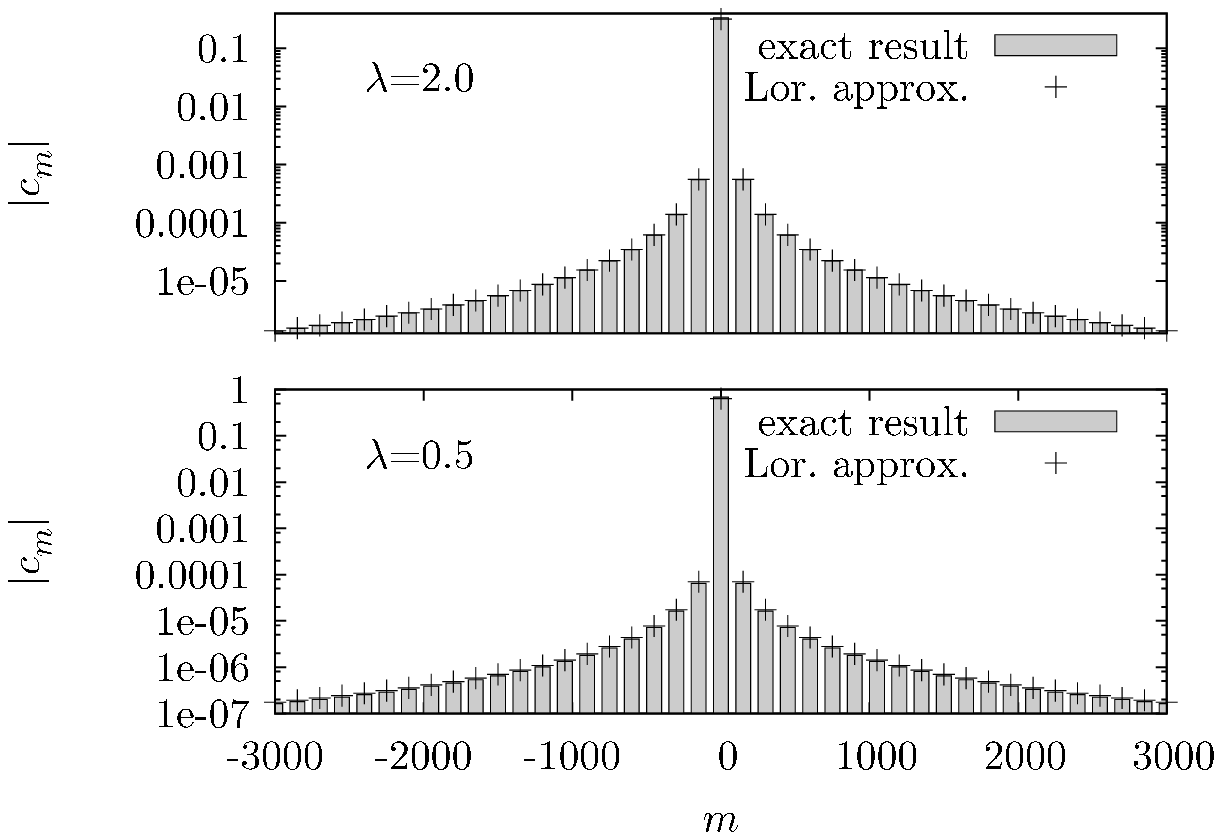}
\caption{\label{fig:loroamdist} Orbital angular momentum distribution for 
different 
values of $\lambda$. a) $\lambda=-2.0$: linear plot shows the distribution
for the central $m$ values. b) $\lambda=-0.5$: linear plot. 
c) $\lambda=-2.0$: logarithmic
plot for the flanks. The width of the bars in the plot covers 100 $m$ values,
but the shown value corresponds to the $m$ value in the middle of the bar.
d) $\lambda=-0.5$: logarithmic plot.}
\end{center}
\end{figure}

The exact angular momentum distribution has a shape similar to a
Lorentzian (see figure \ref{fig:oamdist}).
We therefore seek to
approximate the Fourier kernel as an exponentially decaying function. Owing to
the symmetry of the wavefunction the Fourier cosine transform can be used:
\begin{equation}
\label{eq:fouriertrafo}
c_m = \frac{2}{N\sqrt{2\pi}} \int_{-\pi}^0 \exp(\frac{|\lambda|}{2} 
\phi^2) \cos(m \phi) \rmd\phi.
\end{equation}
In essence the integrand can be approximated in the same way as for 
the wavefunction approximation.
The Fourier integrand can be written as $\exp[|\lambda|(\phi+\pi)(\phi-\pi)/2]
\exp(|\lambda|\pi^2/2)$. Within the integration interval $-\pi \leq \phi <0$ 
and for 
large $|\lambda|$ this Fourier kernel is only significantly different from zero
around the lower boundary $-\pi$. This allows us to approximate the
kernel with $\exp[|\lambda|(\phi+\pi)(-\pi)]\exp(|\lambda|\pi^2/2)$ and
to extend the integration interval in (\ref{eq:fouriertrafo}) to
infinity, which results in an angular momentum distribution in the shape of a 
Lorentzian:
\begin{eqnarray}
\label{eq:lorcm}
c_m & \approx & \frac{1}{N\sqrt{2\pi}} \exp(\frac{|\lambda|}{2}\pi^2)
\int_{-\pi}^\infty  \exp(-\pi|\lambda|(\phi+\pi))
\cos(m\phi)\rmd \phi, \nonumber\\
& = & \frac{1}{N\sqrt{2\pi}} (-1)^m \exp(\frac{|\lambda|}{2}\pi^2)
\left( \frac{2|\lambda|\pi}{|\lambda|^2\pi^2 + m^2} \right).
\end{eqnarray}
The Lorentzian and wavefunction approximation are very similar but not 
identical. This is because 
extending the upper integration boundary is not an equally good approximation for the
two integrals in (\ref{eq:lorcm}) and (\ref{eq:wvfnorm}). 
One can see from figure (\ref{fig:loroamdist}) that the agreement with the 
Lorentzian is better at the flanks and for higher values of $\lambda$. This
is consistent with our considerations as for larger values of $|\lambda|$ the
justification for the approximation of the wavefunction in the Fourier 
integral becomes
more valid. Also, for central values of $m$ the region around the 
upper boundary in the Fourier integral (\ref{eq:fouriertrafo}) contributes
more.
The angular momentum probabilities have to sum to unity. This condition
can be used to calculate the normalisation constant within the 
Lorentzian approximation. We denote this approximate normalisation
constant with $N_\mathrm{Lor}$ to distinguish it from the analytically
exact value $N$. Using the results for the probability amplitudes $c_m$ in 
(\ref{eq:lorcm}) the sum of the probabilities may be written as:
\begin{equation}
\sum_{m=-\infty}^\infty | c_m |^2 = 1 = \frac{1}{N_{\mathrm{Lor}}^2 2\pi} 
4 |\lambda|^2 \pi^2 \exp(|\lambda|\pi^2) \sum_{m=-\infty}^\infty 
\left( |\lambda|^2\pi^2 + m^2 \right)^{-2}.
\end{equation}
The summation can be executed using contour integration 
\cite{stephenson+:cup:1993}, which results in the following expression
for $N_\mathrm{Lor}$:
\begin{equation}
\label{eq:lornorm}
N_{\mathrm{Lor}}^2 = \exp(|\lambda|\pi^2) \left( \pi 
\cosech^2(|\lambda| \pi^2 ) + \frac{1}{|\lambda|\pi} \coth(|\lambda|\pi^2) 
\right).
\end{equation}

The angular momentum uncertainty can also be calculated from the 
angular momentum probabilities $|c_m|^2$ in the Lorentzian
approximation (\ref{eq:lorcm}) in the same way as the normalisation
constant $N_{\mathrm{Lor}}$. From figure (\ref{fig:loroamdist}) and
equation (\ref{eq:lorcm}) it can be seen that the angular momentum mean
is zero. The square of the uncertainty is thus given by:
\begin{equation}
(\Delta m)^2 \approx \frac{1}{N_{\mathrm{Lor}}^2 2\pi} 
4 |\lambda|^2 \pi^2 \exp(|\lambda|\pi^2) \sum_{m=-\infty}^\infty 
\frac{m^2}{\left(|\lambda|^2\pi^2 + m^2 \right)^2}.
\end{equation}
Using again contour integration to evaluate the sum 
\cite{stephenson+:cup:1993} and substituting the value
of $N_{\mathrm{Lor}}$ we can write for the angular momentum
uncertainty in the Lorentzian approximation:
\begin{eqnarray}
\label{eq:lordm}
(\Delta m)^2 & = &  \sum_m |c_m|^2 m^2 \\  
& \approx & \pi^2|\lambda|^2 \frac{ \coth(|\lambda| \pi^2)
 - |\lambda|\pi^2 \cosech^2(|\lambda| \pi^2)}
{\coth(|\lambda| \pi^2) + |\lambda|\pi^2 \cosech^2(|\lambda| \pi^2)}.
\end{eqnarray}
For $|\lambda|\pi^2 > 3$ this expression simplifies because the hyperbolic
function $\cosech(|\lambda|\pi^2)$ becomes negligible small. In this limit
we have
\begin{equation}
(\Delta m) \approx \pi|\lambda|.
\end{equation}
The behaviour of the angular momentum uncertainty as a function of $\lambda$
can be seen in figure (\ref{fig:oamuncertainty}).

%%%
%%% Conclusions
%%%

\section{Conclusion}
We have studied the uncertainty relation for angular momentum and angle in
a series of recent papers. The progress in creating and manipulating
optical states with orbital angular momentum made it possible to 
confirm the form of the angular uncertainty relation experimentally for
intelligent states, that is states which obey the equality in the
uncertainty relation \cite{frankearnold+:njp6:2004}. 
As the angular uncertainty relation has a state dependent lower bound, the
intelligent states do not necessarily minimize the uncertainty product. We have
explored the difference between the intelligent states and states, which give a
minimum in the uncertainty product under additional constraints in a second
paper \cite{pegg+:njp7:2005}. In the present paper we emphasize the difference
between intelligent states and minimum product states by introducing
a class of intelligent states with arbitrarily large uncertainty product.

The difference between the two kinds of intelligent states is given by 
the sign of a parameter $\lambda$. For positive $\lambda$ we have the
truncated Gaussians with an uncertainty product bounded by $\hbar/2$. For 
negative $\lambda$ the wavefunction grows exponentially towards the edge of the
chosen radian range, resulting in an arbitrarily
large uncertainty product. We have 
derived analytical expressions for the angular momentum and angle uncertainties and
the uncertainty product. In two limiting cases, for small negative $\lambda$
and large negative $\lambda$ we have explained the behaviour of the uncertainties
with the parameter $\lambda$ using approximate results.
The three papers combined give an extensive overview of the uncertainty relation for
angular momentum and angle.

%%% 
%%% Acknowledgements
%%%

\ack
We would like to thank David T Pegg for helpful discussions and we acknowledge
financial support from the UK Engineering and Physical Sciences Research Council
(EPSRC) under the grant GR S03898/01 and the Royal Society of Edinburgh.

%%%
%%% References
%%%

%%%
%%% End of document
%%%

\end{document}